\documentclass[12pt]{iopart}

\usepackage{iopams}
\usepackage{epsfig}

\begin{document}

\title{Quantum games with decoherence}
\author{A P Flitney and D Abbott}
\address{Centre for Biomedical Engineering and \\
Department of Electrical and Electronic Engineering, \\
The University of Adelaide, Adelaide, SA 5005, Australia}
\ead{aflitney@eleceng.adelaide.edu.au, dabbott@eleceng.adelaide.edu.au}

\begin{abstract}
A protocol for considering decoherence in quantum games is presented.
Results for two-player, two-strategy quantum games
subject to decoherence are derived
and some specific examples are given.
Decoherence in other types of quantum games is also considered.
As expected, the advantage that a quantum player achieves over a player
restricted to classical strategies is diminished for increasing decoherence
but only vanishes in the limit of maximum decoherence.
\end{abstract}

\submitto{\JPA}
\maketitle

\section{Introduction}
\label{sec-intro}
Game theory has long been commonly used in economics,
the social sciences and biology to model decision making situations
where the outcomes are contingent upon the interacting strategies
of two or more agents with conflicting, or at best, self-interested
motives.
There is now increasing interest in applying game-theoretic techniques
in physics~\cite{abbott02}.
With the enthusiasm for quantum computation there has been a surge
of interest in the discipline of quantum information~\cite{nielsen00}
that has lead to the creation of a new field
combining game theory and quantum mechanics:
quantum game theory~\cite{meyer99}.
By replacing classical probabilities with quantum amplitudes
and allowing the players to employ
superposition, entanglement and interference,
quantum game theory produces new ideas
from classical two-player
\cite{eisert99,eisert00,marinatto00,iqbal01,iqbal02,flitney02a}
and multi-player settings~\cite{benjamin01,kay01,du02a,flitney04}.
Quantum prisoners' dilemma has been realized on a two qubit
nuclear magnetic resonance machine~\cite{du02b}.
A review of quantum games is given by Flitney and Abbott~\cite{flitney02b}.

Decoherence can be defined as non-unitary dynamics resulting
from the coupling of the system with the environment.
In any realistic quantum computer,
interaction with the environment cannot be entirely eliminated.
Such interaction can destroy the special features of quantum computation.
A recent review of the standard mechanisms of quantum decoherence can be found in reference~\cite{zurek03}. 
Quantum computing in the presence of noise is possible with the use
of quantum error correction~\cite{preskill98}
or decoherence free subspaces~\cite{lidar03}.
These techniques work by encoding the
logical qubits in a number of physical qubits.
Quantum error correction is successful, provided the error rate is low enough,
while decoherence free subspaces control certain types of decoherence.
Both have the disadvantage of expanding the number
of qubits required for a calculation.
Without such measures, the theory of  quantum control in the presence
of noise and decoherence is little studied.
This motivates the study of quantum games, which can be viewed as a game-theoretic
approach to quantum control
--- game-theoretic methods in classical control theory~\cite{carraro95}
are well-established and translating them to the quantum realm
is a promising area of study.
Johnson has considered a quantum game corrupted by noisy input~\cite{johnson01}.
Above a certain level of noise it was found that the quantum effects
impede the players to such a degree
that they were better off playing the classical game.
Chen {\em et al} have discussed decoherence
in quantum prisoners' dilemma~\cite{chen03}.
Decoherence was found to have no effect on the Nash equilibrium
in this model.
The current work considers general quantum games
in the presence of decoherence.
The paper is organized as follows.
Section~\ref{sec-games} outlines our model for introducing
decoherence into quantum games,
section~\ref{sec-results} presents some specific results from this model
for two-player, two-strategy quantum games,
section~\ref{sec-other} gives an example of decoherence in another quantum game
and section~\ref{sec-conc} presents concluding remarks.

\section{Quantum games with decoherence}
\label{sec-games}
The process of quantizing a game with two pure strategies proceeds as follows.
In the classical game the possible actions of a player
can be encoded by a bit.
This is replaced by a qubit in the quantum case.
The computational basis states $|0\rangle$ and $|1\rangle$ represent
the classical pure strategies,
with the players' qubits initially prepared in the $|0\rangle$ state.
The players' moves are unitary operators,
or more generally, completely positive, trace-preserving maps,
drawn from a set of strategies $S$,
acting on their qubits.
Interaction between the players' qubits is necessary
for the quantum game to give something new.
Eisert {\em et al} produced
interesting new features by introducing entanglement~\cite{eisert99}.
The final state of an $N$-player quantum game in this model is computed by
\begin{equation}
\label{e-qgame}
| \psi_f \rangle = \hat{J}^{\dagger} (\hat{M}_1 \otimes \hat{M}_2
	\otimes \ldots \otimes \hat{M}_N) \, \hat{J} | \psi_0 \rangle \;,
\end{equation}
where
$|\psi_0 \rangle = |00 \ldots 0\rangle$ represents the initial state of the $N$ qubits,
$\hat{J}$ ($\hat{J^{\dagger}}$) is an operator
that entangles (dis-entangles) the players' qubits,
and $\hat{M}_k, \; k = 1,\ldots,N$, represents the move of player $k$.
A measurement over the computational basis is taken on $|\psi_f\rangle$
and the payoffs are subsequently determined
using the payoff matrix of the classical game.
The two classical pure strategies are the identity
and the bit flip operator.
The classical game is made a subset of the quantum one
by requiring that $\hat{J}$ commute
with the direct product of $N$ classical moves.
Games with more than two classical pure strategies are catered for by
replacing the qubits by qunits ($n$ level quantum systems)
or, equivalently, by associating with each player a number of qubits.
For a discussion of the formalism of quantum games see~\cite{lee02}.

It is most convenient to use the density matrix notation
for the state of the system
and the operator sum representation for the quantum operators.
Decoherence can take many forms including dephasing,
which randomizes the relative phases of the quantum states,
and dissipation that modifies the populations of the quantum states.
Pure dephasing of a qubit can be expressed as
\begin{equation}
        a |0\rangle \:+\: b |1\rangle
                \;\rightarrow\; a |0\rangle \:+\: b \, e^{\rmi \phi} |1\rangle.
\end{equation}
If we assume that the phase kick $\phi$ is a random variable
with a Gaussian distribution of mean zero and variance $2 \lambda$,
then the density matrix obtained after averaging over all values of $\phi$ is~\cite{nielsen00}
\begin{equation}
	\left( \begin{array}{cc}
                |a|^2 & a \bar{b} \\
                \bar{a} b & |b|^2
	\end{array} \right)
        \;\rightarrow\; \left( \begin{array}{cc}
                |a|^2 & a \bar{b} \, e^{-\lambda} \\
                \bar{a} b \, e^{-\lambda} & |b|^2
	\end{array} \right).
\end{equation}
Over time, the random phase kicks cause an exponential decay
of the off-diagonal elements of the density matrix.

In this work we shall use the quantum operator formalism to model decoherence.
This method is well known to have its limitations~\cite{royer96}.
For a good description of the quantum operator formalism
and an example of its limitations the reader is referred to chapter 8 of reference~\cite{nielsen00}.
Other methods for calculating decoherence include using
Lagrangian field theory, path integrals, master equations, quantum Langevin equations,
short-time perturbation expansions, Monte-Carlo methods, semiclassical methods,
and phenomenological methods~\cite{brandt98}.

In the operator sum representation,
the act of making a measurement with probability $p$
in the $\{|0\rangle, |1\rangle \}$ basis
on a qubit $\rho$ is
\begin{equation}
\rho \;\rightarrow\; \sum_{j=0}^{2} {\cal E}_j \, \rho \, {\cal E}_j^{\dagger},
\end{equation}
where ${\cal E}_0 = \sqrt{p} |0\rangle \langle 0|$,
${\cal E}_1 = \sqrt{p} |1\rangle \langle 1|$
and ${\cal E}_2 = \sqrt{1-p} \, \hat{I}$.
An extension to $N$ qubits is achieved by applying the measurement
to each qubit in turn, resulting in
\begin{equation}
\label{e-measure}
\rho \;\rightarrow\; \sum_{j_1, \ldots, j_N = 0}^{2} {\cal E}_{j_1} \otimes \ldots \otimes {\cal E}_{j_N}
		\, \rho \, {\cal E}_{j_N}^{\dagger} \otimes \ldots \otimes {\cal E}_{j_1}^{\dagger},
\end{equation}
where $\rho$ is the density matrix of the $N$ qubit system.
This process also leads to the decay of the off-diagonal elements of $\rho$.
By identifying $1-p = e^{-\lambda}$,
the measurement process has the same results as pure dephasing.

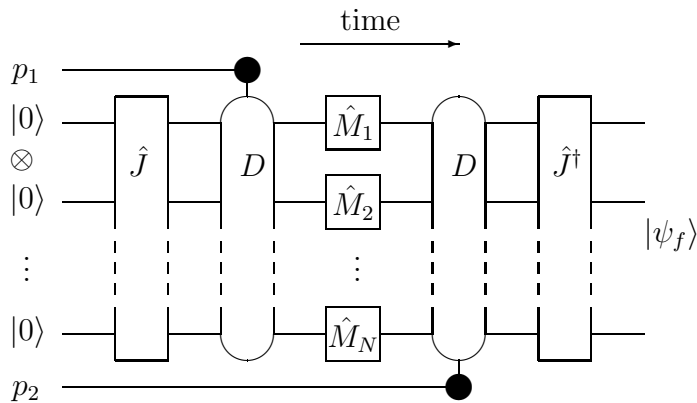
\begin{figure}
\begin{center}
\begin{picture}(260,140)(0,-10)
	\multiput(0,58)(0,30){2}{$|0\rangle$}
	\put(0,8){$|0\rangle$}
	\put(0,73){$\otimes$}
	\put(240,45){$|\psi_f\rangle$}

	\multiput(20,60)(40,0){6}{\line(1,0){20}}
	\multiput(20,90)(40,0){6}{\line(1,0){20}}
	\put(45,70){$\hat{J}$}
	\put(205,70){$\hat{J}^{\dagger}$}

	\multiput(40,50)(20,0){2}{\line(0,1){50}}
	\multiput(200,50)(20,0){2}{\line(0,1){50}}
	\multiput(40,100)(160,0){2}{\line(1,0){20}}

	\put(120,80){\framebox(20,20){$\hat{M}_1$}}
	\put(120,50){\framebox(20,20){$\hat{M}_2$}}
	\put(120,0){\framebox(20,20){$\hat{M}_N$}}
	\multiput(5,30)(125,0){2}{$\vdots$}

	\multiput(40,0)(160,0){2}{\line(1,0){20}}
	\multiput(40,0)(20,0){2}{\line(0,1){20}}
	\multiput(200,0)(20,0){2}{\line(0,1){20}}
	\multiput(20,10)(40,0){6}{\line(1,0){20}}

	\put(90,50){\oval(20,100)[t]}
	\put(90,20){\oval(20,40)[b]}
	\put(170,50){\oval(20,100)[t]}
	\put(170,20){\oval(20,40)[b]}
	\multiput(87,70)(80,0){2}{$D$}

	\multiput(40,25)(20,0){4}{\line(0,1){4}}
	\multiput(40,33)(20,0){4}{\line(0,1){4}}
	\multiput(40,41)(20,0){4}{\line(0,1){4}}
	\multiput(160,25)(20,0){4}{\line(0,1){4}}
	\multiput(160,33)(20,0){4}{\line(0,1){4}}
	\multiput(160,41)(20,0){4}{\line(0,1){4}}

	\put(90,110){\circle*{10}}
	\put(170,-10){\circle*{10}}

	\put(90,105){\line(0,-1){5}}
	\put(170,-5){\line(0,1){5}}
	\put(20,110){\line(1,0){65}}
	\put(20,-10){\line(1,0){145}}

	\put(1,107){$p_1$}
	\put(1,-13){$p_2$}

	\put(110,120){\vector(1,0){60}}
	\put(125,125){time}

\end{picture}
\end{center}
\caption{The flow of information in an $N$-person quantum game
with decoherence,
where $M_k$ is the move of the $k$th player
and $\hat{J}$ ($\hat{J}^{\dagger}$) is an entangling (dis-entangling) gate.
The central horizontal lines are the players' qubits and
the top and bottom lines are classical random bits with a probability
$p_1$ or $p_2$, respectively, of being 1.
Here, $D$ is some form of decoherence
controlled by the classical bits.}
\label{f-qgame}
\end{figure}

Independently of the particular model used,
a quantum game with decoherence can be described in the following manner
\begin{eqnarray}
\label{e-scheme}
	\rho_i \equiv & \rho_0 = | \psi_0 \rangle \langle \psi_0 |
			& {\rm (initial \; state)} \nonumber\\
	& \rho_1 = \hat{J} \rho_0 \hat{J}^{\dagger} & \mbox{(entanglement)} \nonumber\\
	& \rho_2 = D(\rho_1, p_1)
		& \mbox{(partial decoherence)} \nonumber\\
	& \rho_3 = (\otimes_{k=1}^{N} \hat{M}_{k}) \, \rho_2 \,
			(\otimes_{k=1}^{N} \hat{M}_{k})^{\dagger}
		& \mbox{(players' moves)} \nonumber\\
	& \rho_4 = D(\rho_3, p_2)
		& \mbox{(partial decoherence)} \nonumber\\
	& \rho_5 = \hat{J}^{\dagger} \rho_4 \hat{J} &
		\mbox{(dis-entanglement)},
\end{eqnarray}
to produce the final state $\rho_f \equiv \rho_5$ upon which a measurement is taken.
The function $D(\rho, p)$ is a completely positive map
that applies some form of decoherence to the state $\rho$
controlled by the probability $p$.
The scheme is shown in figure~\ref{f-qgame}.
The expectation value of the payoff for the $k$th player is
\begin{equation}
	\langle \$^k \rangle =
		\sum_{\alpha} \hat{P}_{\alpha} \, \rho_f
			\, \hat{P}_{\alpha}^{\dagger} \, \$_{\alpha}^{k},
\end{equation}
where $P_{\alpha} = |\alpha \rangle \langle \alpha|$
is the projector onto the state $|\alpha \rangle$,
$\$_{\alpha}^{k}$ is the payoff to the $k$th player
when the final state is $|\alpha\rangle$,
and the summation is taken over 
$\alpha = j_1 j_2 \ldots j_N, \; j_i = 0,1$.

\section{Results for $2 \times 2$ quantum games}
\label{sec-results}
Let $S = \{ \hat{U}(\theta, \alpha, \beta): \: 0 \le \theta \le \pi, \: -\pi \le \alpha,\beta \le \pi \}$
be the set of pure quantum strategies, where
\begin{equation}
\label{e-qstrategy}
\hat{U}(\theta, \alpha, \beta) =
	\left( \begin{array}{cc}
		e^{\rmi \alpha} \cos (\theta/2) & \rmi e^{\rmi \beta} \sin (\theta/2) \\
		i e^{-\rmi \beta} \sin (\theta/2) & e^{-\rmi \alpha} \cos(\theta/2)
	    \end{array} \right)
\end{equation}
is an SU(2) operator.
The move of the $k$th player is $\hat{U}(\theta_k, \alpha_k, \beta_k)$.
The classical moves are $\hat{I} \equiv \hat{U}(0,0,0)$ and $\hat{F} \equiv \hat{U}(\pi,0,0)$.
Entanglement is achieved by~\cite{benjamin01}
\begin{equation}
\hat{J} = \frac{1}{\sqrt{2}} (\hat{I}^{\otimes N} \,+\: \rmi \sigma_x^{\otimes N}).
\end{equation}
Operators from the set $S_{\rm cl} = \{ \hat{U}(\theta,0,0): \: 0 \le \theta \le \pi \}$
are equivalent to classical mixed strategies since,
when all players use these strategies,
the quantum game reduces to the classical one.
There is some arbitrariness about the representation of the operators.
Different representations will only lead to a different overall phase in the final state
and this has no physical significance.

After choosing equation~(\ref{e-measure}) to represent the function $D$ in (\ref{e-scheme}),
we are now in a position to write down the results of decoherence
in a $2 \times 2$ quantum game.
Using the subscripts $A$ and $B$ to indicate the parameters
of the two traditional protagonists Alice and Bob, respectively,
and writing $c_k \equiv \cos(\theta_k/2)$ and $s_k \equiv \sin(\theta_k/2)$
for $k={\scriptstyle A,B}$,
the expectation value of a player's payoff is
\begin{eqnarray}
\fl \langle \$ \rangle
	=\frac{1}{2}(c_{\rm A}^2 c_{\rm B}^2 + s_{\rm A}^2 s_{\rm B}^2)(\$_{00} + \$_{11})
		\:+\: \frac{1}{2}(c_{\rm A}^2 s_{\rm B}^2 + s_{\rm A}^2 c_{\rm B}^2)(\$_{01} + \$_{10})
\nonumber\\
	+ \frac{1}{2} (1-p_1)^2 (1-p_2)^2 \{ [c_{\rm A}^2 c_{\rm B}^2 \cos (2 \alpha_{\rm A} + 2 \alpha_{\rm B})
			\,-\, s_{\rm A}^2 s_{\rm B}^2 \cos (2 \beta_{\rm A} + 2 \beta_{\rm B})] (\$_{00} - \$_{11})
\nonumber\\
	\quad + [c_{\rm A}^2 s_{\rm B}^2 \cos (2 \alpha_{\rm A} - 2 \beta_{\rm B})
			\,-\, s_{\rm A}^2 c_{\rm B}^2 \cos (2 \alpha_{\rm B} - 2 \beta_{\rm A})] (\$_{01} - \$_{10}) \}
\nonumber\\
	+ \frac{1}{4} \sin \theta_{\rm A} \sin \theta_{\rm B} \,
		\left[ (1-p_1)^2 \sin(\alpha_{\rm A} + \alpha_{\rm B} - \beta_{\rm A} - \beta_{\rm B})
			(-\$_{00} + \$_{01} + \$_{10} - \$_{11}) \right.
\nonumber\\
	\quad + (1-p_2)^2 \sin(\alpha_{\rm A} + \alpha_{\rm B} + \beta_{\rm A} + \beta_{\rm B})
				(\$_{00} - \$_{11})
\nonumber\\
	\quad \left. + (1-p_2)^2 \sin(\alpha_{\rm A} - \alpha_{\rm B} + \beta_{\rm A} - \beta_{\rm B})
				(\$_{10} - \$_{01}) \right],
\end{eqnarray}
where $\$_{ij}$ is the payoff to the player for the final state $|ij\rangle$.
Setting $p_1 = p_2 = 0$ gives the quantum games
of the Eisert {\em et al} model~\cite{eisert99}
studied in the literature.
If in addition, $\alpha_k = \beta_k = 0, \; k={\scriptstyle {\rm A,B}}$,
a $2 \times 2$ classical game results
with the mixing between the two classical pure strategies $\hat{I}$ and $\hat{F}$
being determined by $\theta_{\rm A}$ and $\theta_{\rm B}$ for Alice and Bob, respectively.
Maximum decoherence with $p_1 = p_2 = 1$
gives a result where the quantum phases $\alpha_k$ and $\beta_k$ are not relevant:
\begin{equation}
\langle \$ \rangle = \frac{x}{2}(\$_{00} + \$_{11}) \:+\: \frac{1-x}{2}(\$_{01} + \$_{10}),
\end{equation}
where $x = c_{\rm A}^2 c_{\rm B}^2 + s_{\rm A}^2 s_{\rm B}^2$.
In a symmetric game the payoff to both players is the same
and the game is not equivalent to the original classical game.
Extrema for the payoffs occur when both $\theta$'s are 0 or $\pi$.

One way of measuring the ``quantum-ness'' of the game is to consider the known advantage
of a player having access to the full set of quantum strategies $S$
over a player who is limited to the classical set $S_{\rm cl}$~\cite{eisert99,flitney03}.
If we restrict Alice to $\alpha_{\rm A} = \beta_{\rm A} = 0$, then,
\begin{eqnarray}
\label{e-payqc}
\fl \langle \$ \rangle
	= \frac{x}{2}(\$_{00} + \$_{11}) \:+\: \frac{1-x}{2}(\$_{01} + \$_{10})
\nonumber\\
	+ \frac{1}{2} (1-p_1)^2 (1-p_2)^2 \left\{ c_{\rm B}^2 \cos 2 \alpha_{\rm B} \,
		[ c_{\rm A}^2 (\$_{00} - \$_{11}) + s_{\rm A}^2 (\$_{10} - \$_{01}) ] \right.
\nonumber\\
	\quad \left. - s_{\rm B}^2 \cos 2 \beta_{\rm B} \,
		[ c_{\rm A}^2 (\$_{10} - \$_{01}) + s_{\rm A}^2 (\$_{00} - \$_{11}) ] \right\}
\nonumber\\
	+ \frac{1}{4} \sin \theta_{\rm A} \sin \theta_{\rm B}
		\left[ (1-p_1)^2 \sin(\alpha_{\rm B} - \beta_{\rm B})
			(-\$_{00} + \$_{01} + \$_{10} - \$_{11}) \right.
\nonumber\\
	\quad \left. + (1-p_2)^2 \sin(\alpha_{\rm B} + \beta_{\rm B})
			(\$_{00} + \$_{01} - \$_{10} - \$_{11}) \right].
\end{eqnarray}

\begin{figure}
\begin{center}
\epsfig{file=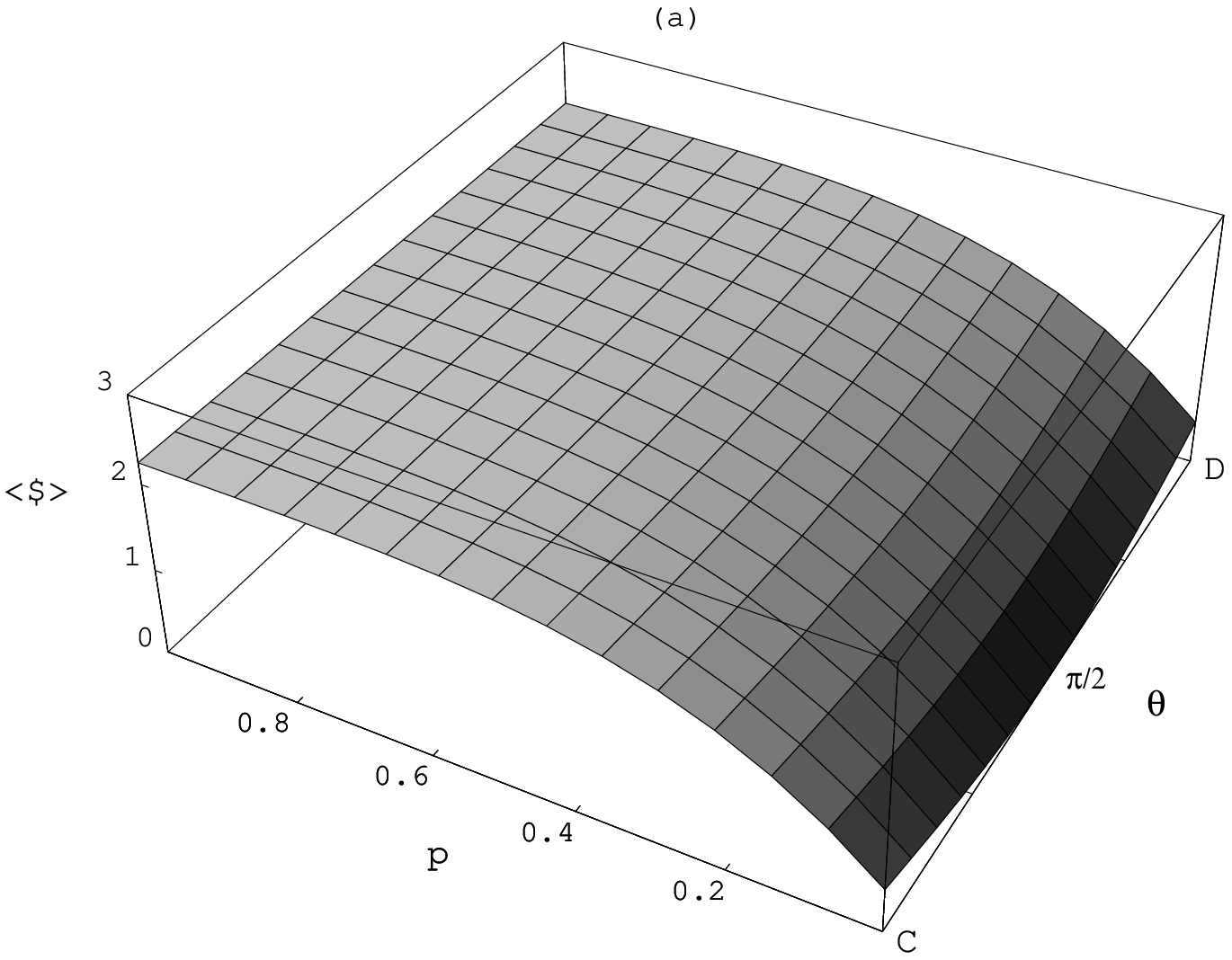, width=7cm}
\epsfig{file=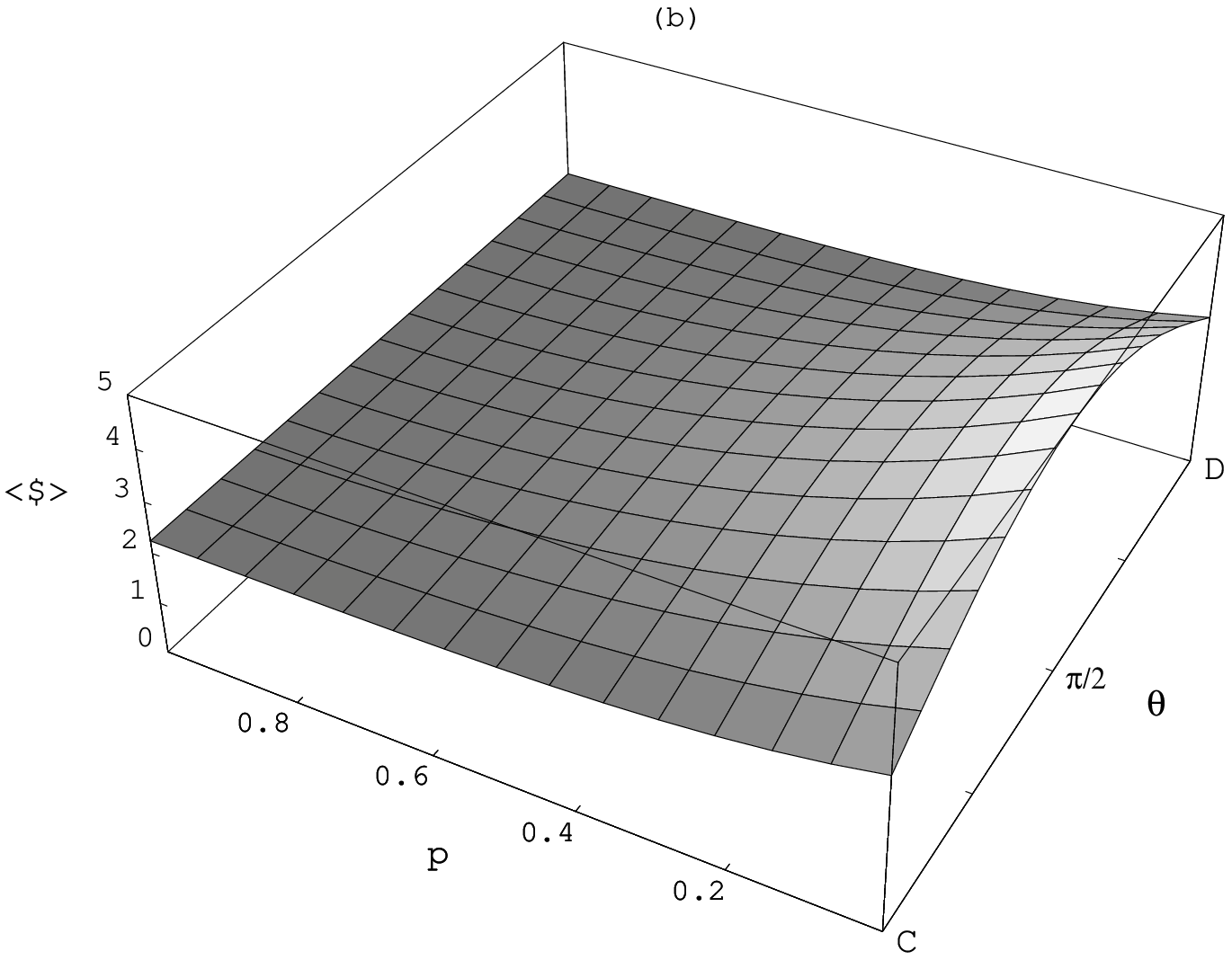, width=7cm}
\end{center}
\caption{Payoffs for (a) Alice and (b) Bob in quantum prisoners' dilemma
as a function of decoherence probability $p$ and Alice's strategy $\theta$
(being a measure of the mixing between cooperation ($C$) and defection ($D$)
with $\theta=0$ giving $C$ and $\theta=\pi$ giving $D$),
when Bob plays the optimum quantum strategy and Alice is restricted to classical strategies.
The decoherence goes from the unperturbed quantum game at $p=0$
to maximum decoherence at $p=1$.}
\label{f-pd} 
\end{figure}

For prisoners' dilemma, the standard payoff matrix is
\begin{equation}
\label{e-pd}
  \begin{array}{c|cc}
	\mbox{prisoners'} & \mbox{Bob :} \\
	\mbox{dilemma} & \mbox{cooperation}(C) & \mbox{defection}(D) \\
	\hline
	\makebox[2cm]{Alice :} C & (3,3) & (0,5) \\
	\makebox[2cm]{} D & (5,0) & (1,1)
  \end{array}
\end{equation}
where the numbers in parentheses represent payoffs to Alice
and Bob, respectively.
The classical pure strategies are cooperation ($C$) and defection ($D$).
Defecting gives a better payoff regardless of the other player's strategy,
so it is a dominant strategy,
and mutual defection is the Nash equilibrium.
The well known dilemma arises from the fact that both players
would be better off with mutual cooperation,
if this could be engineered.
With the payoffs of equation~(\ref{e-pd}),
the best Bob can do from equation~(\ref{e-payqc}) is to select
$\alpha_{\rm B} = \pi/2$ and $\beta_{\rm B} = 0$.
Bob's choice of $\theta_{\rm B}$ will depend on Alice's choice of $\theta_{\rm A}$.
He can do no better than $\theta_{\rm B} = \pi/2$
if he is ignorant of Alice's strategy\footnote{See
Flitney and Abbott~\cite{flitney03} for details
of quantum versus classical players.}.
Figure~\ref{f-pd} shows Alice and Bob's payoffs
as a function of decoherence probability $p \equiv p_1 = p_2$
and Alice's strategy $\theta \equiv \theta_{\rm A}$
when Bob selects this optimal strategy.

\begin{figure}
\begin{center}
\epsfig{file=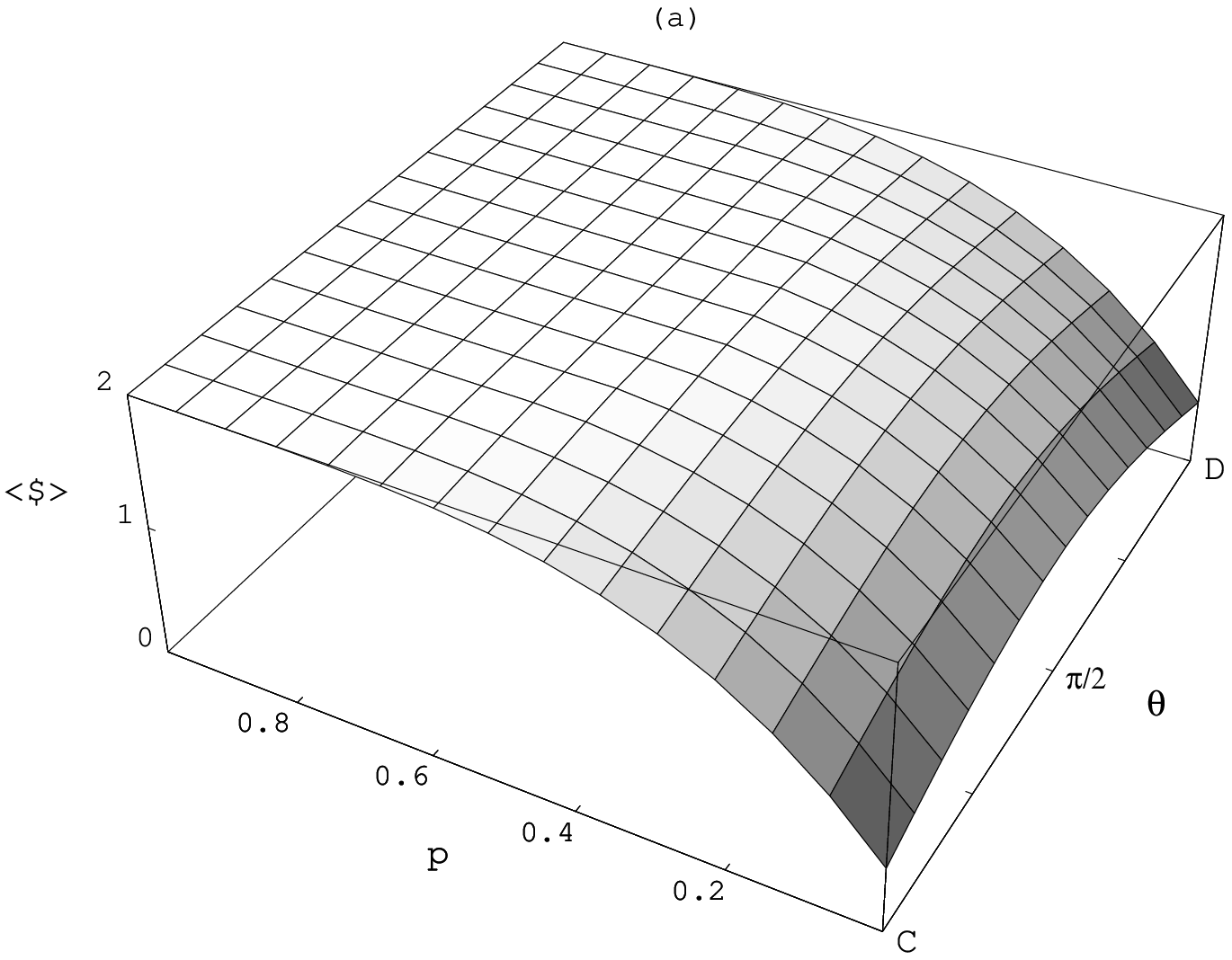, width=7cm}
\epsfig{file=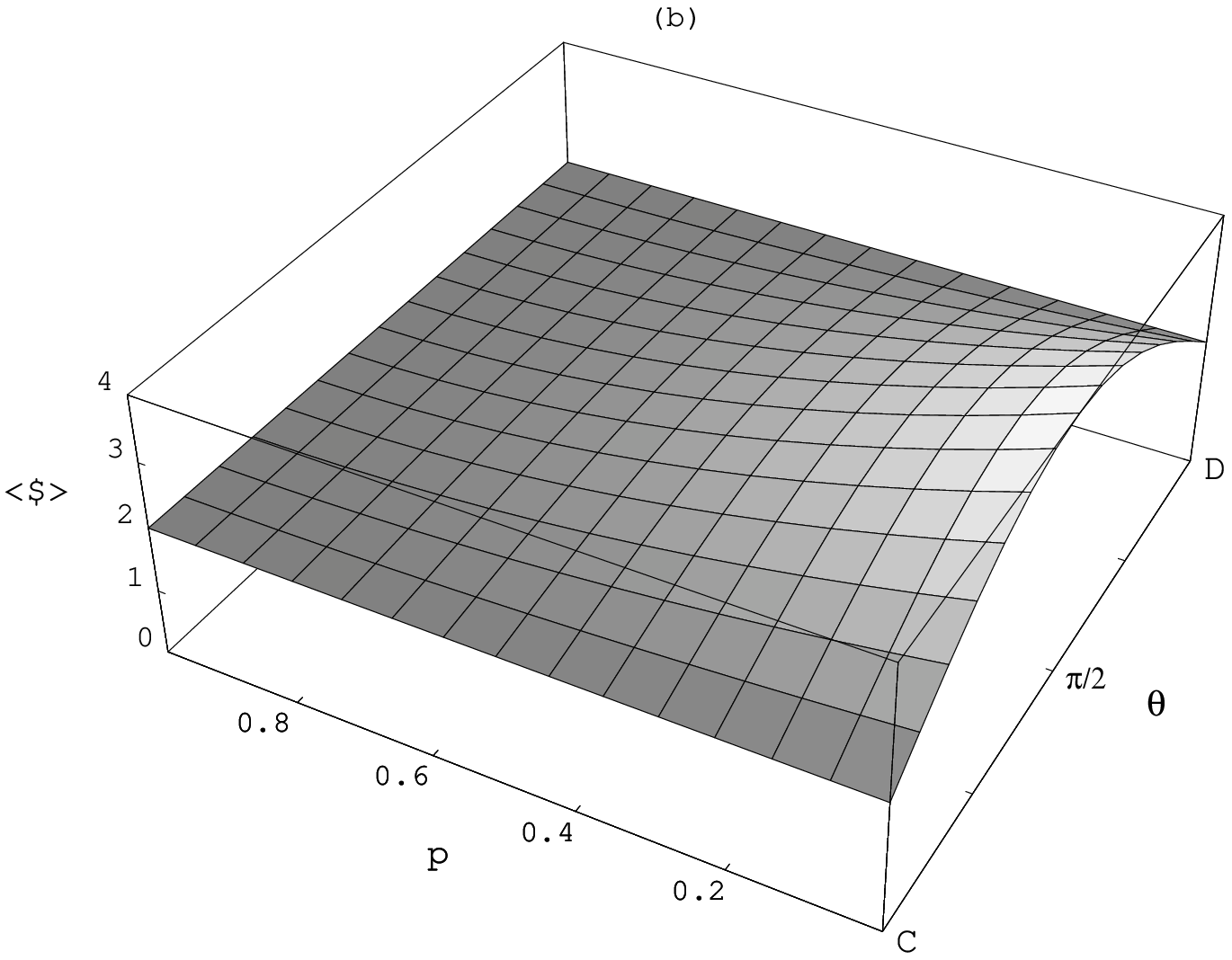, width=7cm}
\end{center}
\caption{Payoffs for (a) Alice and (b) Bob in quantum chicken
as a function of decoherence probability $p$ and Alice's strategy $\theta$,
when Bob plays the optimum quantum strategy and Alice is restricted to a classical
mixed strategy.}
\label{f-chicken} 
\end{figure}

The standard payoff matrix for the game of chicken is
\begin{equation}
\label{e-chicken}
  \begin{array}{c|cc}
	 & \mbox{Bob :} \\
	\mbox{chicken} & \mbox{cooperation}(C) & \mbox{defection}(D) \\
	\hline
	\makebox[2cm]{Alice :} C & (3,3) & (1,4) \\
	\makebox[2cm]{} D & (4,1) & (0,0)
  \end{array}
\end{equation}
There is no dominant strategy.
Both $CD$ and $DC$ are Nash equilibria,
with the former preferred by Bob and the latter by Alice.
Again there is a dilemma since the Pareto optimal results $CC$
is different from both Nash equilibria.
As above, Bob's payoff is optimized by $\alpha_{\rm B} = \pi/2$, $\beta_{\rm B} = 0$
and $\theta_{\rm B} = \pi/2$.
Figure~\ref{f-chicken} shows the payoffs
as a function of decoherence probability $p$
and Alice's strategy $\theta$.

\begin{figure}
\begin{center}
\epsfig{file=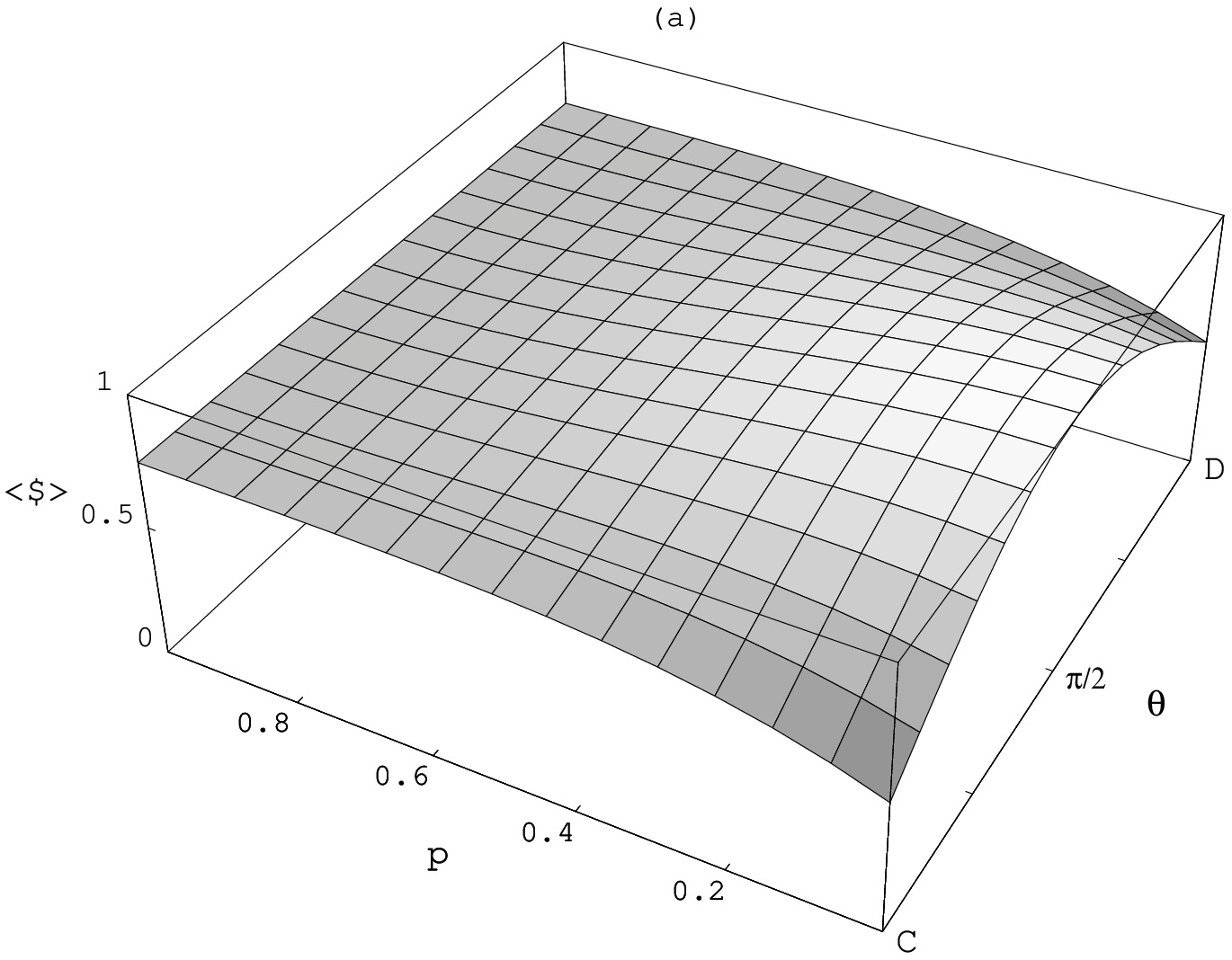, width=7cm}
\epsfig{file=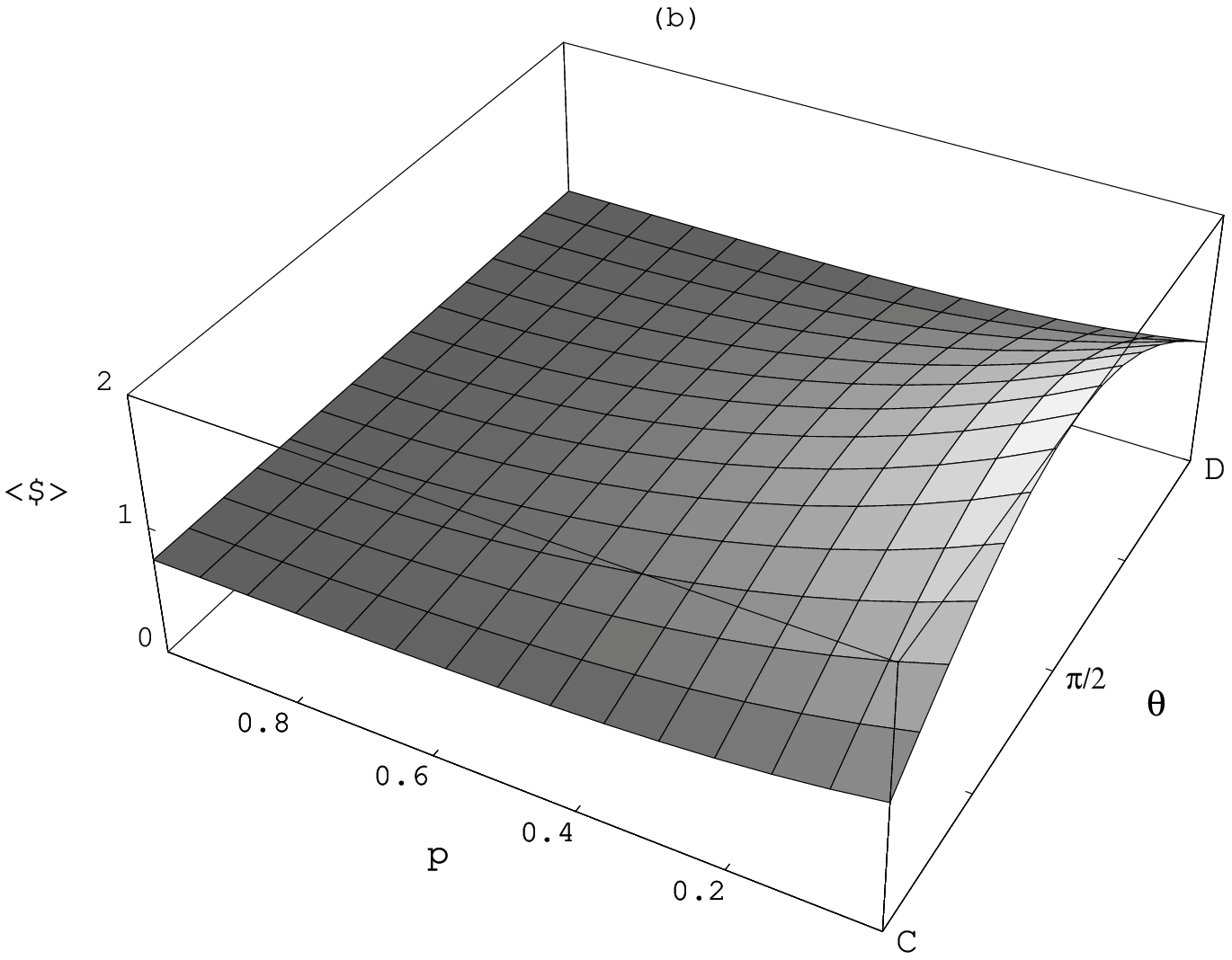, width=7cm}
\end{center}
\caption{Payoffs for (a) Alice and (b) Bob in quantum battle of the sexes
as a function of decoherence probability $p$ and Alice's strategy $\theta$,
when Bob plays the optimum quantum strategy and Alice is restricted to a classical
mixed strategy.}
\label{f-bos}
\end{figure}

One form of the payoff matrix for the battle of the sexes is
\begin{equation}
\label{e-bos}
  \begin{array}{c|cc}
	\mbox{battle} & \mbox{Bob :} \\
	\mbox{of the sexes} & \mbox{opera}(O) & \mbox{television}(T) \\
	\hline
	\makebox[2cm]{Alice :} O & (2,1) & (0,0) \\
	\makebox[2cm]{} T & (0,0) & (1,2)
  \end{array}
\end{equation}
Here the two protagonist must decide on an evening's entertainment.
Alice prefers opera ($O$) and Bob television ($T$),
but their primary concern is that they do an activity together.
In the absence of communication there is a coordination problem.
A quantum Bob maximizes his payoff in a competition with a classical Alice
by choosing $\alpha_{\rm B} = -\pi/2$, $\beta_{\rm B} = 0$ and $\theta_{\rm B} = \pi/2$.
Figure~\ref{f-bos} shows the resulting payoffs for Alice and Bob
as a function of decoherence probability $p$
and Alice's strategy $\theta$.

\begin{figure}
\begin{center}
\epsfig{file=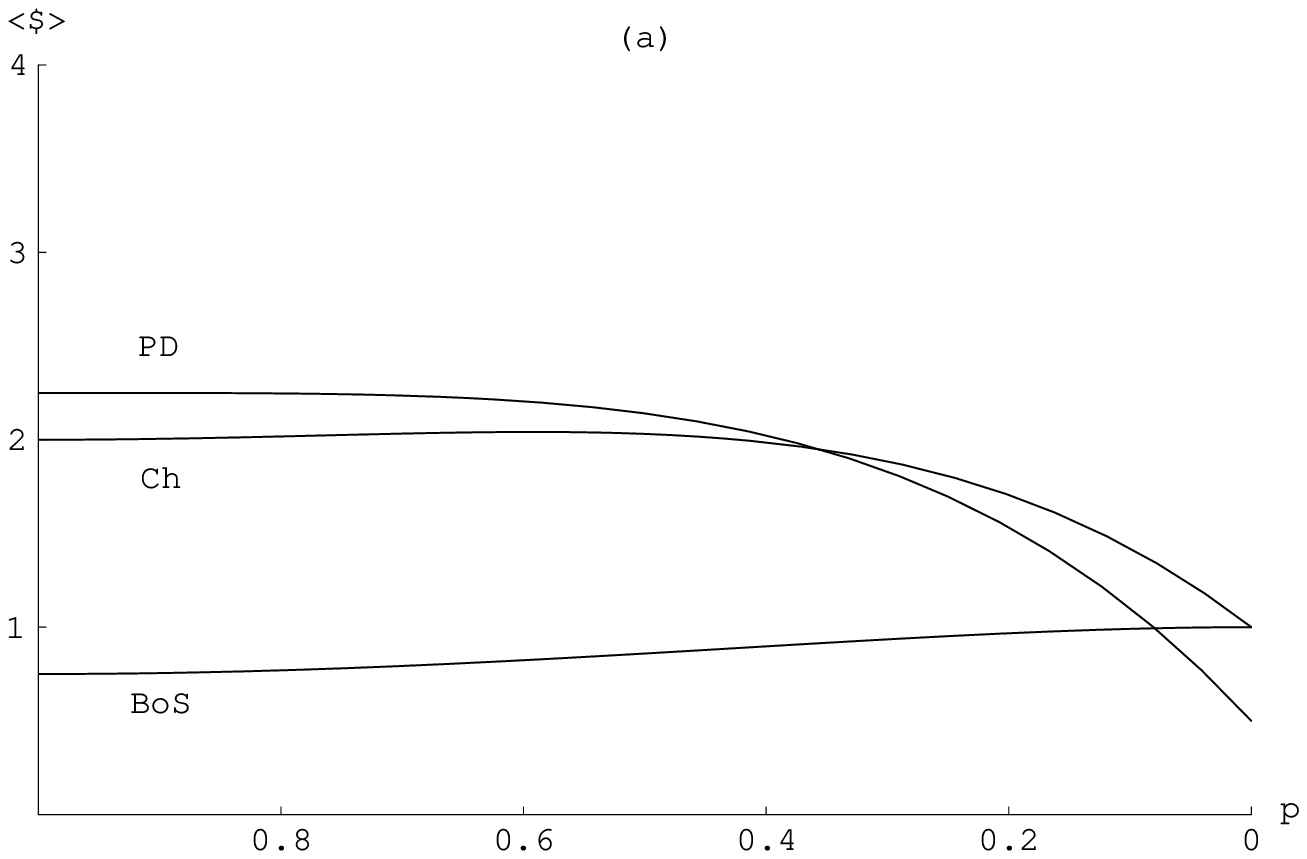, width=7cm}
\epsfig{file=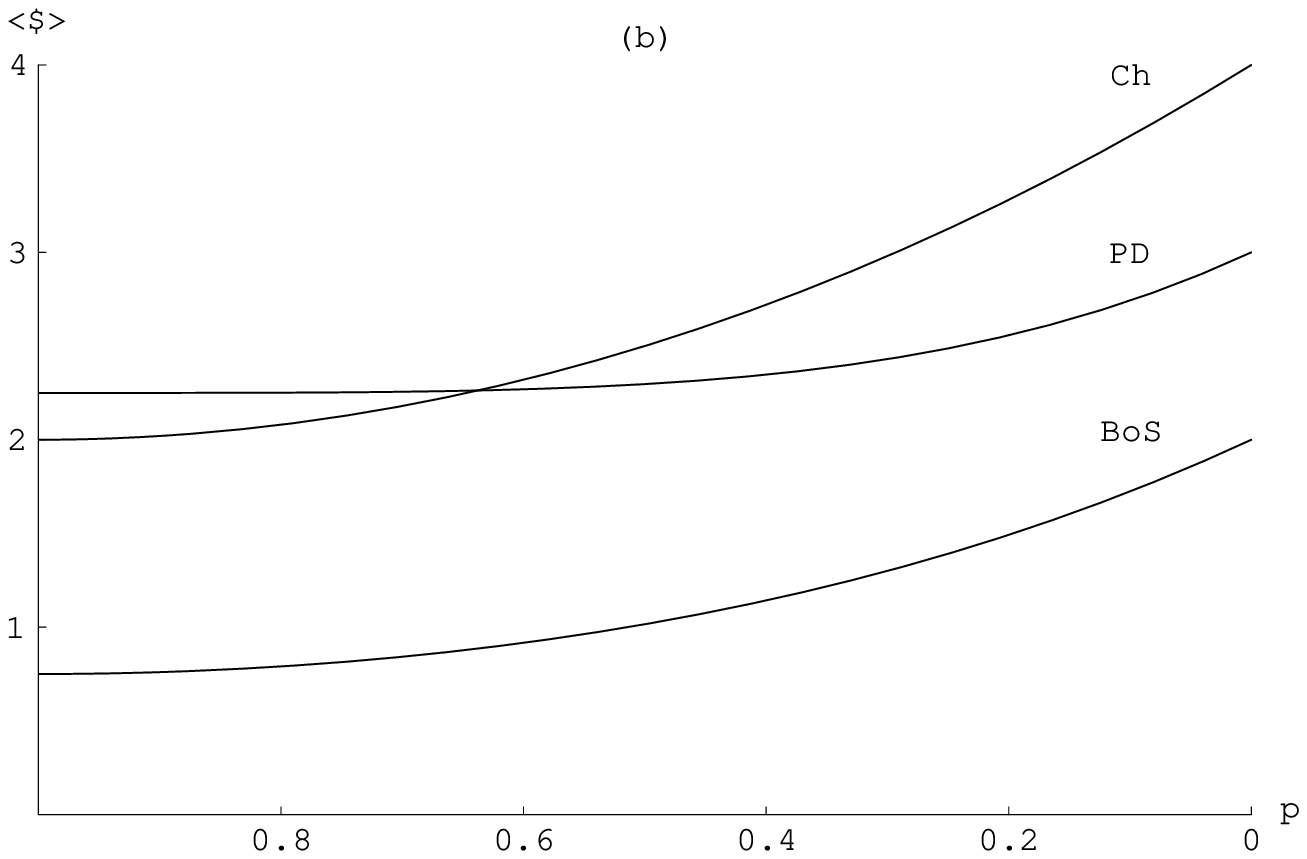, width=7cm}
\end{center}
\caption{Payoffs as a function of decoherence probability $p$,
going from fully decohered on the left ($p=1$)
to fully coherent on the right ($p=0$),
for (a) Alice and (b) Bob for the quantum games
prisoners' dilemma (PD), chicken (Ch) and battle of the sexes (BoS).
Bob plays the optimum quantum strategy and Alice her best classical
counter strategy.
As expected, the payoff to the quantum player, Bob,
increases with increasing coherence
while Alice performs worse except in the case of battle of the sexes.
This game is a coordination game 
--- both players do better if they select the same move ---
and Bob can increasingly engineer coordination as coherence improves,
helping Alice as well as himself.}
\label{f-best}
\end{figure}

The optimal strategy for Alice in the three games considered
is $\theta = \pi$ (or 0) for prisoners' dilemma,
or $\theta = \pi/2$ for chicken and battle of the sexes.
Figure~\ref{f-best} shows the expectation value of the payoffs
to Alice and Bob as a function of the decoherence probability $p$
for each of the games
when Alice chooses her optimal classical strategy.
In all cases considered,
Bob outscores Alice and performs better than his classical Nash equilibrium result
provided $p<1$.\footnote{or the poorer of his two Nash equilibria
in the case of chicken or the battle of the sexes}
The advantage of having access to quantum strategies
decreases as $p$ increases,
being minimal above $p \approx 0.5$,
but is still present for all levels of decoherence up to the maximum.
At maximum decoherence ($p=1$),
with the selected strategies,
the game result is randomized
and the expectation of the payoffs
are simply the average over the four possible results.
The results presented in figures~\ref{f-pd}, \ref{f-chicken} and \ref{f-bos}
are comparable to the results for different levels of entanglement~\cite{flitney03}.
They are also consistent with the results of Chen and co-workers~\cite{chen03}
who show that with increasing decoherence the payoffs to both players
approach the average of the four payoffs in a quantum prisoners' dilemma.

\section{Decoherence in other quantum games}
\label{sec-other}
A simple effect of decoherence can be seen
in Meyer's quantum penny-flip~\cite{meyer99}
between P, who is restricted to classical strategies,
and Q, who has access to quantum operations.
In the classical game,
P places a coin heads up in a box.
First Q, then P, then Q again,
have the option of (secretly) flipping the coin or leaving it unaltered,
after which the state of the coin is revealed.
If the coin shows heads, Q is victorious.
Since the players' moves are carried out in secret
they do not know the intermediate states of the coin
and hence the classical game is balanced.

In the quantum version,
the coin is replaced by a qubit prepared
in the $|0\rangle$ (``heads'') state.
Having access to quantum operations,
Q applies the Hadamard operator
to produce the superposition $(|0\rangle + |1\rangle)/\sqrt{2}$.
This state is invariant under
the transformation $|0\rangle \leftrightarrow |1\rangle$
so P's action has no effect.
On his second move
Q again applies the Hadamard operator to return the qubit to $|0\rangle$.
Thus Q wins with certainty against any classical strategy by P.

Decoherence can be added to this model
by applying a measurement with probability $p$
after Q's first move.
Applying the same operation after P's move
has the same effect since his move
is either the identity or a bit-flip.
If the initial state of the coin is represented by the density matrix
$\rho_0 = |0\rangle \langle 0|$,
the final state can be calculated by
\begin{equation}
\eqalign{
\rho_f	&= \hat{H} \hat{P} \hat{D} \hat{H} \rho_0
		\hat{H}^{\dagger} \hat{D}^{\dagger} \hat{P}^{\dagger} \hat{H}^{\dagger} \\
	&= \frac{1}{4}
	   \left( \begin{array}{cc}
		4 - 2 p & 0 \\
		0 & 2p
	   \end{array} \right),
}
\end{equation}
where $\hat{H}$ is the Hadamard operator,
$\hat{P}$ is P's move ($\hat{I}$ or $\sigma_x$), and
$\hat{D} = \sqrt{1-p} \, \hat{I} \:+\: \sqrt{p}(|0\rangle \langle 0| \,+\, |1\rangle \langle 1|)$
is a measurement in the computational basis with probability $p$.
Again, the final state is independent of P's move.
The expectation of Q winning
decreases linearly from one to $\frac{1}{2}$
as $p$ goes from zero to one.
Maximum decoherence produces a fair game.

As an example of the effect of decoherence on another quantum game
consider a game analogous to a three player duel, or truel,
between Alice, Bob and Charles~\cite{flitney04}.
The classical version can be described as follows.
Each player has a bit, starting in the one state.
The players move in sequence in alphabetic order.
A move consists of either doing nothing or attempting to flip an opponent's bit
with a known probability of failure of $a$, $b$ or $c$,
for Alice, Bob and Charles, respectively.
A player can do nothing if their bit is zero.
The payoffs at the completion of the game are 1/(number of bits in the one state)
to a player whose bit is one,
or zero otherwise.
(The connection with a truel is made by considering one
to correspond to ``alive'' and zero to ``dead''.
A move is an attempt to shoot an opponent.)
In some situation the optimal strategy is counter-intuitive.
It may be beneficial for a player to do nothing
rather than attempt to flip an opponent's bit from one to zero,
since if they are successful they become the target for the third player.

\begin{figure}
\begin{center}
\epsfig{file=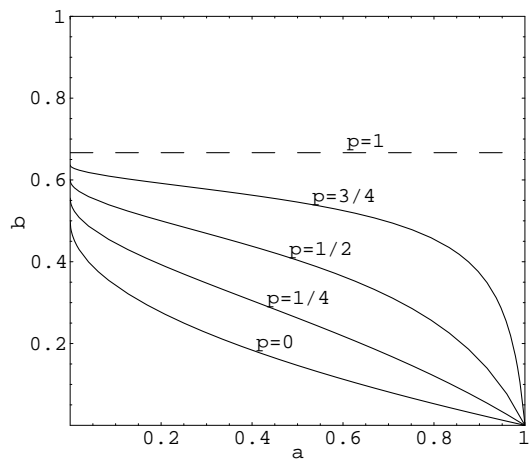, width=7cm}
\end{center}
\caption{In a one round quantum truel with $c=0$ and with decoherence,
the boundaries for different values of the decoherence probability $p$
below which Alice maximizes her expected payoff by doing nothing
and above which by targeting Charles.
There is a smooth transition from the fully quantum case $(p=0)$
to the classical one $(p=1)$.
From Flitney and Abbott~\cite{flitney04}.}
\label{f-truel}
\end{figure}

The game is quantized by replacing the players' bits by qubits
and by replacing the flip operation by an SU(2) operator of the form of
equation~(\ref{e-qstrategy})
operating on the chosen qubit.
Maintaining coherence throughout the game removes the dynamic aspect
since the players can get no information on the success of previous moves.
Noise can be added to the quantum game
by giving a probability $p$ of a measurement being made
after each move,
and in the case of a measurement,
allowing the players to choose their strategy
depending on the result of previous rounds,
which are now known.
Figure~\ref{f-truel} shows the regions of the parameter space $(a,b)$
corresponding to Alice's preferred strategy
in a one round truel when $c=0$
(i.e., when Charles is always successful).
The boundary between Alice maximizing her expected payoff by doing nothing
and by targeting Charles depends on the decoherence
probability $p$.
We see a smooth transition from the quantum case
to the classical as $p$ goes from zero to one.
Note that the boundary in the parameter space changes from linear in the classical case
to convex in the quantum case.
This is of interest since convexity is being intensely studied
as the basis for Parrondo's paradox~\cite{harmer02,harmer04}
and the current example may provide an opportunity
for generating a quantum Parrondo's paradox~\cite{meyer02,flitney02c,ng04}.

\section{Conclusion}
\label{sec-conc}
A method of introducing decoherence into quantum games has been presented.
One measure of the ``quantum-ness'' of a quantum game subject to decoherence
is the advantage a quantum player has over a player restricted to classical strategies.
As expected,
increasing the amount of decoherence degrades the advantage of the quantum player.
However, in the model considered,
this advantage does not entirely disappear
until the decoherence is a maximum.
When this occurs in a $2 \times 2$ symmetric game,
the results of the players are equal.
The classical game is not reproduced.
The loss of advantage to the quantum player
is very similar to that which occurs
when the level of entanglement between the players' qubits is reduced.

In the example of a one-round quantum truel,
increasing the level of decoherence altered the regions of parameter space
corresponding to different preferred strategies smoothly toward the classical regions.
In this quantum game, maximum decoherence produces a situation
identical to the classical game.

In multi-player quantum games
it is known that new Nash equilibria can arise~\cite{benjamin01}.
The effect of decoherence on the existence
of the new equilibria is an interesting open question.
There has been some work on continuous-variable quantum games~\cite{li02}
involving an infinite dimensional Hilbert space.
The study of decoherence in infinite dimensional Hilbert space quantum games
would need to go beyond the simple quantum operator method presented in this paper
and is yet to be considered.

This paper has focused on {\it static} quantum games and so future work
on game-theoretic methods for dynamic quantum systems with different
types of decohering noise will be of great interest.
A particular open question will be to compare the behavior of such quantum games for
(a) the non-Markovian case, where the quantum system is coupled to a
dissipative environment with memory,
with (b) the Markovian (memoryless) limit where the correlation times,
in the decohering environment,
are small compared to the characteristic time scale of the quantum system.

\ack      
This work was supported by GTECH Corporation Australia
with the assistance of the SA Lotteries Commission (Australia).
We thank Jens Eisert of Potsdam University for useful discussions.

\end{document}